\documentclass[a4paper,11pt]{article}
\pdfoutput=1
\usepackage{jinstpub}
\usepackage{lineno}

\usepackage{subcaption}
\usepackage[capitalise,nameinlink]{cleveref} 
\usepackage{color}

\title{\boldmath A flexible test facility for liquid xenon detector development}

\author[a]{Evan~Angelico,}
\author[a,1]{Jacopo~Dalmasson\note{Now at: Fondazione Bruno Kessler, Italy.},}
\author[a]{Ralph~DeVoe,}
\author[a,b]{Giorgio~Gratta,}
\author[a,*]{Clarke~A.~Hardy\note[*]{Corresponding author.},}
\author[a,2]{Brian~Lenardo\note{Now at: SLAC National Accelerator Laboratory, Menlo Park, CA, USA.},}
\author[a]{Lin~Si,}
\author[a]{Marie~Vidal,}
\author[a,3]{and Shuoxing~Wu\note{Now at: Fermi National Accelerator Laboratory, Batavia, IL, USA.}}
\affiliation[a]{Physics Department, Stanford University, Stanford, CA 94305, USA}
\affiliation[b]{Hansen Experimental Physics Lab, Stanford University, Stanford, CA 94305, USA}
\emailAdd{cahardy@stanford.edu}

\abstract{As liquid xenon time projection chambers scale to ever-larger sizes, so too do the engineering challenges they pose. We describe a large, flexible, multipurpose test facility capable of supporting the development of a number of key aspects of liquid xenon detector systems. Example applications of this facility include characterization of large-area light and charge sensor arrays, tests of xenon purification techniques and materials compatibility, and investigations into high-voltage phenomena. This facility uses an automated and remotely monitored cryo-cooling system based on immersion of the test chamber in a liquid bath rather than conductive coupling, leading to advantages in temperature and pressure stability, as well as increasing required response times in the case of cooling-power loss. Design advantages, operational procedures, and performance of the facility are described, as well as five examples of liquid xenon test chambers that use the facility.}


\begin{document}
\maketitle
\flushbottom

\section{Introduction}
Liquid xenon (LXe) time projection chambers (TPCs) have been established as an essential technology for nuclear and particle physics due to their low intrinsic radioactivity, scalability, and the co-production of scintillation and ionization signals which results in excellent energy resolution and particle identification. Tonne-scale LXe TPCs are currently being used to search for interactions from hypothesized dark matter candidates \cite{aalbers_first_2023,aprile_first_2023,bo_dark_2025}, while planning is underway for a tonne-scale search for neutrinoless double-beta decay ($0\nu\beta\beta$) \cite{adhikari_nexo_2021}. Beyond the near term, even larger detectors \cite{baudis_darwin_2024,aalbers_nextgeneration_2022,avasthi_kilotonscale_2021,abdukerim_pandaxxta_2024} are being studied to push dark matter searches into the ``neutrino fog'' \cite{ohare_new_2021} and extend their sensitivity to $0\nu\beta\beta$.

Continued improvement to LXe TPC technology demands further development of charge and light sensing schemes, as well as amplification and digitization electronics integrated with those sensors, that are adequately scalable and amenable for ultra-low-background operation. Medium-scale test facilities play a critical role in bridging the gap between the single-sensor R\&D scale and full-scale experiments, as simultaneous readout of arrays of sensors poses challenges that may not be apparent in single-sensor R\&D tests. In addition to sensor development, operation of large-scale LXe TPCs requires an improved understanding of high-voltage phenomena in LXe as well as the chemical compatibility of detector materials with LXe, which can compromise charge signal resolution. 

In this paper, we describe a test facility optimized for stable, long-term operation, with the flexibility to accommodate a variety of different test chambers with LXe masses of up to $\sim$~$100~\mathrm{kg}$. Facilities of this kind are well suited to the vast range of R\&D efforts essential to developing large-scale LXe experiments. Similar university and research lab scale facilities exist, including the one described in \cite{baur_xebra_2023}. The Stanford facility has several distinguishing features, illustrated below. 

\section{Overview}
\begin{figure}[htp]
  \centering
  \includegraphics[width=\columnwidth]{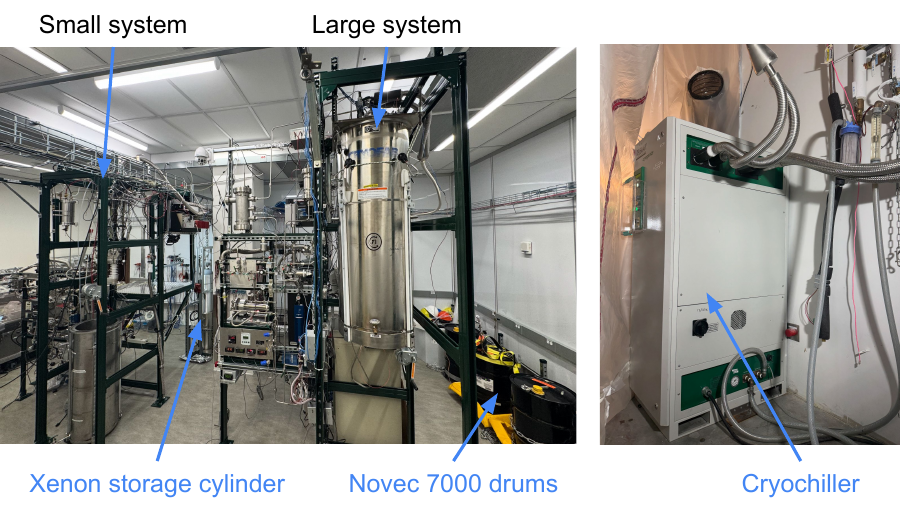}
  \caption{Left: photo of the lab showing the two LXe test stands, the xenon storage cylinder, and the Novec 7000 supply. Right: the cryo chiller outside the lab.}\label{fig:lab}
\end{figure}

The Stanford LXe TPC test facility, shown in \cref{fig:lab}, includes two independent test stands, each with the infrastructure necessary to operate a variety of different LXe test chambers. The ``Large System'' is so named because it uses a cryostat with a 50~cm inner diameter and a 148~cm inner height, while the ``Small System'' uses a cryostat with a 40~cm inner diameter and 118~cm inner height. The Small System and Large System can host test chambers with flange diameters of up to 30.5~cm (DN250CF) and 41.9~cm (DN350CF), respectively. The maximum test chamber height is limited by the space occupied by the inlet and outlet tubing and the housing for sensor electronics above and below the chamber.

Each system is built around a ``top flange'' anchored at the top of a strut channel support frame. The top flange is a steel plate which acts as the feedthrough interface between the air side (above) and the cold space (below). The cold space is contained within a cryostat which can be raised using a rope-and-pulley system to mate with the underside of the top flange, creating a seal with an o-ring. LXe test chambers are suspended in the cryostat from the underside of the top flange using low-thermal-conductivity rods, such as 1/4'' SS316 all-threads or thin-wall SS316 tubing welded to fixture plates. Beside each strut channel frame is an elevated, diamond-tread platform to provide access to the top flange and hold the data acquisition rack, minimizing its distance to the feedthroughs. Both manifolds are connected to the same xenon storage bottle, with valves to direct xenon into and out of either system.

\begin{figure}[htb]
  \centering
  \includegraphics[width=\textwidth]{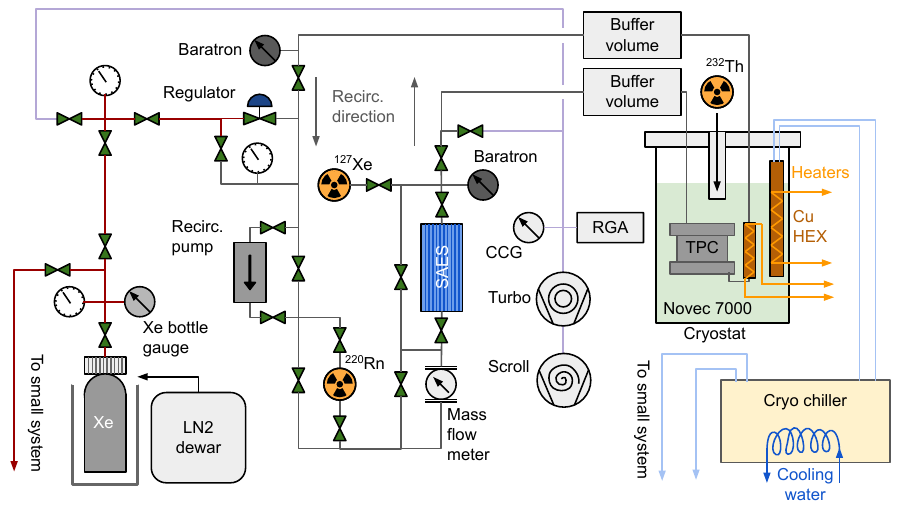}
  \caption{Piping \& instrumentation diagram for the Large System. Red lines represent high-pressure tubing while purple lines represent tubing maintained at vacuum. The Small System shares the xenon supply and the cryo chiller; otherwise the diagram for the Small System is the same.}\label{fig:pid}
\end{figure}

A diagram illustrating one of the xenon systems in its entirety, including xenon manifold, cryostat, and instrumentation, is shown in \cref{fig:pid}. This diagram will be referred to in the following sections related to cooling, xenon handling, and experimental capabilities like radioactive source deployment.

\section{Immersion Cooling}
Unlike similar facilities, which typically use a cold finger conductively coupled to a test chamber in a vacuum space, the systems described here achieve cooling by immersing the test chambers into a temperature-controlled bath of cooling fluid. This alternative technique possesses a number of advantages, particularly with larger-scale detectors. 

The large thermal mass of the cooling fluid, which far exceeds the thermal mass of LXe in the test chambers, results in excellent temperature stability; if cooling is lost due to a failure of the cryostat, the temperature of the LXe rises slowly and in sync with the temperature of the cooling fluid. For experiments using greater than 10~kg LXe masses, which can take multiple days to fill, this level of thermal stability is greatly appreciated by the operators that monitor for failures. 

Furthermore, in the liquid-submerged cryostat, no mechanical thermally-conductive contact is required between the chamber and cooling element. Convective cooling between the liquid, cooling element, and submerged chamber removes the need for a copper LXe chamber or copper braids to achieve thermal uniformity. A chamber of any geometry can be cooled in this fashion, provided it fits inside the cryostat. This allows the LXe chambers to be constructed using off-the-shelf ultra-high-vacuum hardware such as ConFlat (CF), adding convenience and reducing cost.

While only single-phase detectors have been used in these systems, during regular operation all test chambers contain both liquid and gas phases, with the liquid level set high enough to cover all sensors. The liquid level can be set by the mass of xenon filled into the chamber and remains stable throughout a run. For this reason we expect the technique described in this paper to be applicable to both single-phase and dual-phase detectors.

\subsection{Cryogenic System Design}

Cooling power is provided to both systems by a dual-coil, Telemark TVP-2000 cryo chiller, which circulates a proprietary refrigerant in a closed-loop system. The fridge is supplied with room temperature water from the building for cooling its internal heat exchanger. Two sets of vacuum-jacketed refrigerant lines connect the fridge to feedthroughs on the cryostats, allowing for a closed-loop flow of refrigerant through a copper heat exchanger submerged within the cold space of each cryostat. 

Each heat exchanger is suspended from the top flange, and consists of a copper tube with a 6~mm outer diameter brazed in a groove that meanders across a finned copper slab. The copper slabs are shaped as partial cylinders with a diameter close to the inner diameter of the cryostat, allowing for access to the central region where test chambers are suspended. This configuration is designed to promote convection of the cooling fluid. Cartridge heaters are embedded within the copper-slab heat exchangers to increase the rate of warming during xenon recovery. Thermocouples are fixed to the inlet tubing, outlet tubing, and at the top, middle, and bottom of the heat exchanger.

The cooling fluid used in these systems is the hydrofluoroether Novec 7000 produced by 3M \cite{_3m_}. Novec 7000 freezes at 151~K, enabling its use in the 161--175~K temperature range for LXe-based experiments. At 163~K, it has a kinematic viscosity of $8\times10^{-6}~\mathrm{m^2/s}$, providing acceptable convective heat transfer in the LXe temperature range. The use of Novec 7000 as a cooling fluid for LXe TPCs was pioneered by EXO-200 \cite{ackerman_exo200_2022} due to its intrinsic radiopurity, shielding power, viscosity at low temperatures, and thermal mass \cite{leonard_systematic_2008}. 

A typical experiment in the Stanford LXe facility involves filling a cryostat with 200--300~kg of Novec 7000; the substantial thermal mass ensures temperatures at the chamber remain stable, and provides a generous window in which the xenon in the chamber will remain at manageable pressures in the case of a loss of cooling power. Novec 7000 is added by simultaneously cooling the heat exchanger and keeping open a tubing line between the cryostat (at rough vacuum) and a Novec 7000 storage drum. Filling is also facilitated by pressurizing the drum with nitrogen to overcome the hydrostatic column pressure of the fluid in the cryostat. Once the heat exchanger is partially submerged, refrigerator cooling rapidly reduces the Novec 7000 vapor pressure (9.5~psi at room temperature), reducing the pressure required to push liquid from the drum to the cryostat \cite{_3m_}. 

\begin{figure}[htbp]
  \centering
  \includegraphics[width=\columnwidth]{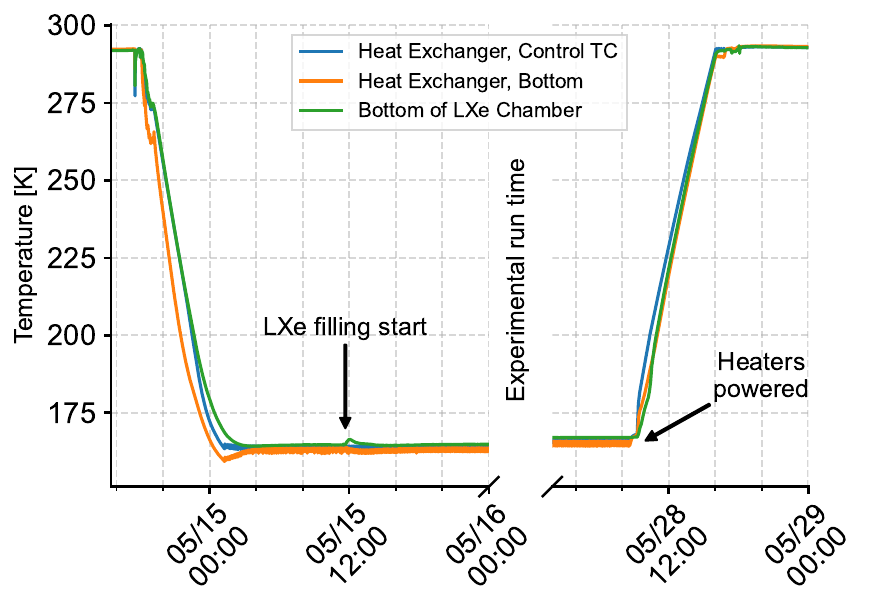}
  \caption{Cooling and warming temperatures and timescales for an experiment in the Large System cryostat. The temperatures during the multi-day experiment are redacted from the x-axis. A visible feature in a thermocouple at the bottom of the test chamber is associated with the time when LXe starts condensing into the chamber.}\label{fig:cooling-and-warming}
\end{figure}

Cooling remains on continuously until the top thermocouple on the heat exchanger reaches the desired filling temperature of 163~K. \cref{fig:cooling-and-warming} shows an example cooling and warming cycle for an experiment in the Large System. Cooling typically takes about 10~hours, though $\sim$5 more hours of normal on-off cooling are required for the temperature distribution throughout the Novec 7000 volume to reach its equilibrium state.

\subsection{Performance}
\begin{figure}[htbp]
  \centering
  \includegraphics[width=\textwidth]{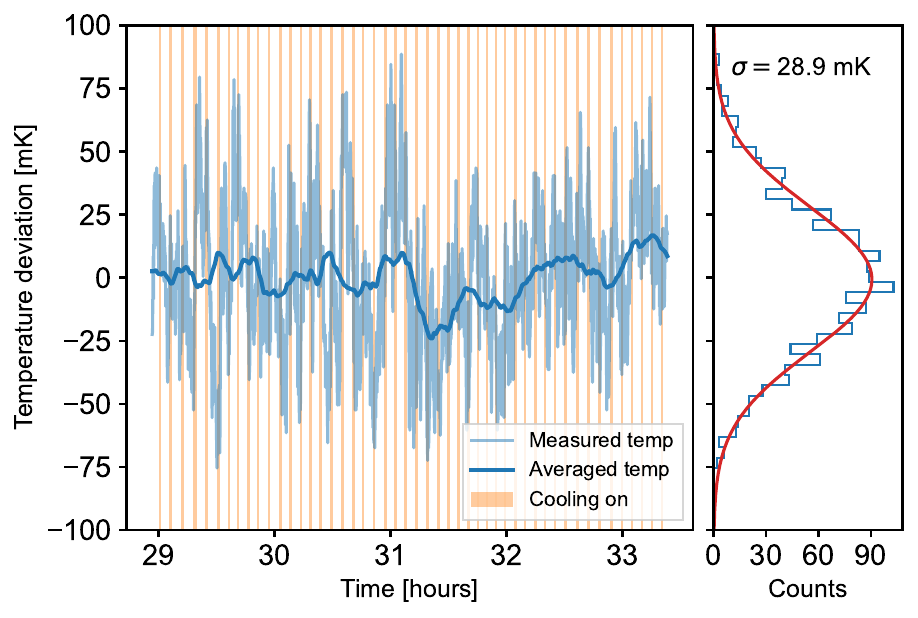}
  \caption{Stability of the test chamber temperature measured by a thermocouple mounted on its top flange. The blue line shows a rolling average computed with a window length of five on-off cooling cycles.}\label{fig:temp}
\end{figure}

During operation, the temperature of a single thermocouple bolted to the heat exchanger is maintained by switching refrigerant flow on or off. A cooling duty cycle of 25\% with a period of 5~minutes is typical. The Novec 7000 bath ensures that the chamber remains insensitive to temperature changes caused by the change in cooling state. Test chamber temperatures have been shown to remain stable to within 100~mK over many hours of operation (\cref{fig:temp} for example). There are no sensors measuring the LXe directly; thermocouples and resistance temperature detectors (RTDs) are instead placed in thermal contact with the exterior surface of the SS316 CF chamber. 

Heat leaking into the cryostat is dominated by conduction through the cryostat top flange --- a solid steel plate, 1.3~cm thick in the Small System and 1.6~cm thick in the Large System --- coupled through gas convection of dry nitrogen between the Novec 7000 liquid surface and the top flange. The nitrogen is used to back-fill the cryostat volume to 1 atmosphere, mitigating potential water vapor leaks from the numerous KF connections and epoxied electrical feedthroughs. Permanent insulation of this region is impractical because of the density of cables and tubes extending through it. 

\begin{figure}[htbp]
  \centering
  \includegraphics[width=\columnwidth]{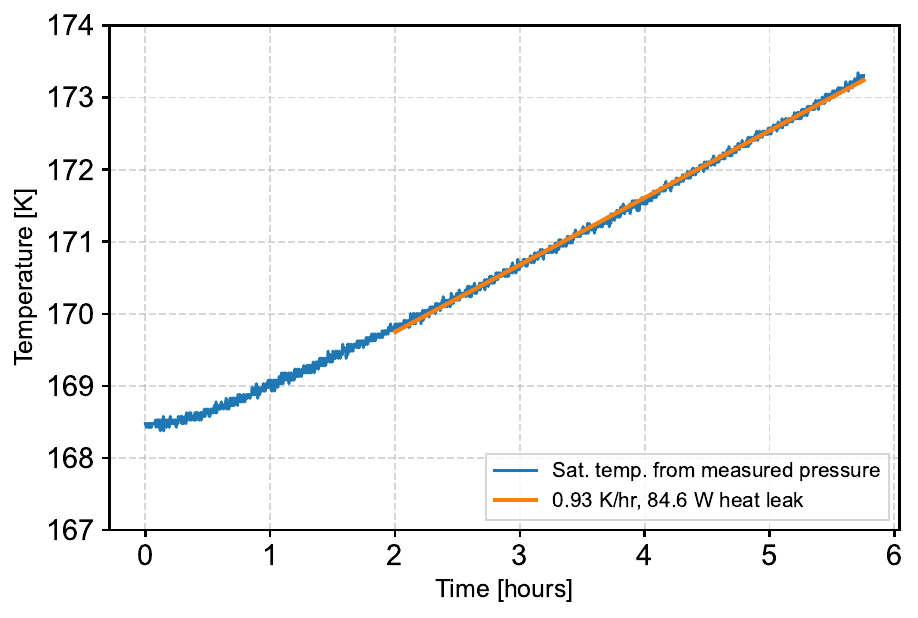}
  \caption{Xenon temperature increase due to natural warming of a test chamber while installed in the Large System. The linear fit is used to estimate an 84.6 W heat leak based on an estimate of the heat capacity of dominant masses in the cryostat. The Novec 7000 mass is 230 kg, the chamber is 38 kg, the copper heat exchanger is 16 kg, and the LXe mass is 9 kg, for a total of capacity of 326 kJ/K. }\label{fig:warming}
\end{figure}

A rate of natural warming of 1~K per hour was measured by stopping cooling intentionally after establishing equilibrium at 168.5~K for multiple days (\cref{fig:warming}). Based on the masses and heat capacities of objects within the cryostat, this implies a roughly 80~W heat leak. Because the LXe systems have been designed with a pressure rating of 1500~torr, operators have approximately 12~hours to remove the xenon in case of a loss of cooling power. While the system status is checked on a more frequent basis than every 12~hours, this flexibility has reduced the burden on operators who otherwise would have had to respond immediately to power outages. See \cref{sec:monitoring} for further detail about the monitoring and alarm system architecture. 

\section{Xenon Handling}
\subsection{Filling \& Recovery}
A single aluminum-body storage cylinder containing $\sim$30~kg of xenon, designed to be submerged in liquid nitrogen (LN2) for cryopumping recovery\footnote{Luxfer SGS N265-SGS}, is used to supply xenon to both systems. This cylinder is permanently plumbed into a high-pressure manifold with valves that can be used to direct flow to one of the two xenon handling manifolds. Each of the handling manifolds uses a clean regulator\footnote{APTech two-stage, tied diaphragm regulators, models AP1720SM and AP1702SM.} to supply xenon to the low-pressure region, while a bypass valve allows flow in the opposite direction during xenon recovery. The xenon storage bottle is suspended from a strain gauge, allowing for real-time monitoring of the mass of xenon that has been filled into an experimental chamber. In addition to the strain gauge, a mass flow meter\footnote{MKS Type 179A and 1479A mass flow meters.} plumbed into the delivery manifold is used to measure xenon flow rates and reconstruct delivered mass. Below the suspended storage bottle is a dewar which can be raised and filled with LN2, submerging the bottom portion of the aluminum storage bottle for use as a xenon cryopump. 

Prior to filling, each system is pumped to vacuum with its own scroll pump and turbomolecular pump (turbopump). Each system is equipped with its own residual gas analyzer (RGA) and cold-cathode gauge (CCG) located near the turbopump. Under typical conditions, the vacuum achievable is $\sim$~$10^{-7}$~hPa at room temperature at the location of the gauge. However, the test chamber is separated from the pump and CCG by $\sim$4 meters of 13~mm diameter tubing, representing an effective conductance of about 0.03~L/s and a source of surface-water outgassing from 1200~cm$^2$ surface area. Lumped-element circuit models are maintained to represent each system's xenon delivery manifold, and are used to estimate pressure as a function of location and infer outgassing rates from control experiments. Depending on the outgassing state inside the chamber and in the manifold tubing, pressures inside the test chambers can fall between 10--100$\times$ larger than the value measured at the pump. 

During filling, the regulator is opened until pressure gauges\footnote{MKS Baratron Capacitance Manometer} at the chamber inlet and outlet read the desired chamber pressure. A heated non-evaporable getter (NEG) purifier\footnote{SAES Monotorr PS3-MT3-R-1.} in the flow path is used to purify the xenon during filling. The filling rate is limited by the rate of condensation at the liquid surface and stainless steel walls, which in turn depends on the surface area in the detector. For this reason, the larger test chambers can be filled at higher flow rates. A typical filling time is $\sim$24~hours for $\sim$30~kg.

Recovery of the liquefied xenon is limited by the temperature of the LXe surface inside the test chamber. To decrease the time it takes to recover LXe, limited by the natural warming rate of 1~K per hour, heaters embedded in the copper heat exchanger apply up to 1 kW of power to the Novec 7000. A constant supply of LN2 is maintained to keep the xenon storage cylinder cold during the recovery process. 

Because the recovery rate exceeds the range of the mass flow meter, and the cylinder mass cannot be measured accurately due to the changing buoyant force from LN2, the pressure in the chamber is used as an indication of the progress of recovery. Early in the recovery process when the xenon is coldest, the flow is throttled using a valve to ensure that cryopumping is not rapid enough to cause the LXe surface to freeze due to evaporative cooling. A finned aluminum heat sink at the storage bottle inlet passively prevents xenon freezing as gas enters and expands in the bottle during fast recovery. Multiple other tube and valve heaters consisting of ceramic heaters embedded in aluminum blocks can be placed around the manifold to prevent freezing at regions most susceptible to cooling by the fast-moving xenon gas. 

As the Novec 7000 surrounding the chamber warms, the valve can be fully opened, at which point the end of the recovery process is indicated with the chamber pressure dropping to below the solidification pressure, solid-xenon vapor pressure, and eventually to < 1 mbar. A typical recovery rate is 10~kg per 3 hours, but can reach as high as 10~kg per hour in cases where the Novec 7000 is pre-warmed prior to recovery.  

\subsection{Recirculation}
\label{sec:recirc}
In both systems, a xenon recirculation pump is used to force flow of xenon gas through a heated NEG purifier to remove electronegative impurities which can capture drifting charges. Both pumps use hermetic volumes in which the only contact with xenon is with stainless steel or teflon. Actuation of a piston draws xenon into the pump inlet and out the pump outlet through one-way valves which set the flow direction.

In the Small System, the pump consists of a large bellows which is driven externally by a pneumatic piston. The Large System uses the pump described in \cite{leport_magnetically_2011,brown_magneticallycoupled_2018}, with a magnetically-coupled piston driven via a stepper motor coupled to a linear translation stage. The controller for the stepper motor can run customized programs for different flow conditions, depending on the pressure, impedance, and flow rate.

\begin{figure}[htbp]
  \centering
  \includegraphics[width=\textwidth]{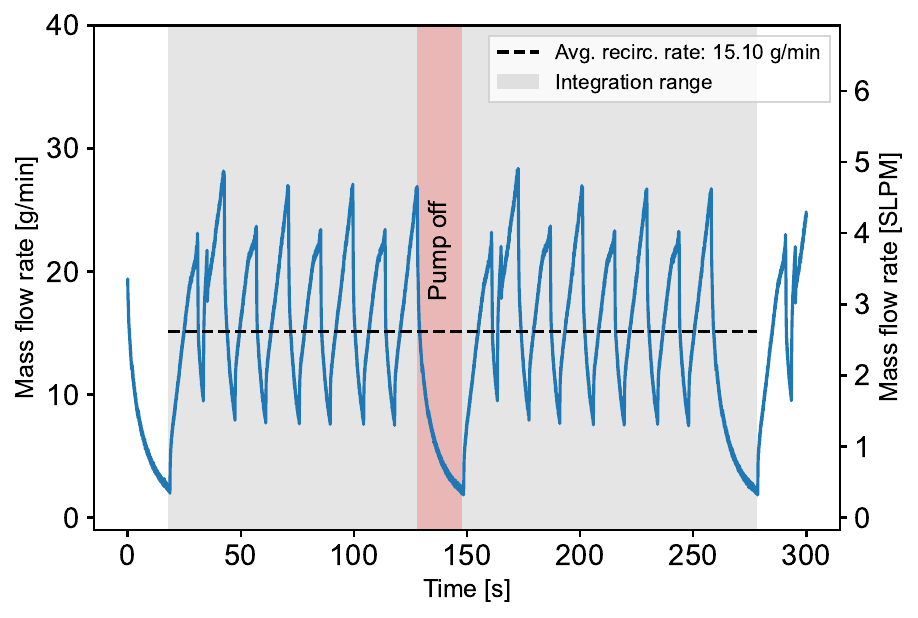}
  \caption{Mass flow rate measured during recirculation through a heated NEG purifier. Each linear stroke of the pump creates a positive pressure differential, but at the turning points the flow decreases, creating the periodic triangular shape shown above. The pump is turned off for a short period of time to allow the xenon gas at the inlet to condense at high pressures, resulting in a duty cycle that is typically greater than 80\% (the off period is shown in shaded red). The average flow rate is calculated by integrating two periods of this cycle (shaded grey).}
  \label{fig:recirc}
\end{figure}

Flow rate readings in both systems are continually recorded, allowing for a measurement of the time to recirculate the entire mass of xenon contained within the chamber, which is a key parameter in determining equilibrium with impurities. An example of instantaneous mass flow rate over a 5 minute period of time using the Large System xenon pump is shown in \cref{fig:recirc}. Due to both the rapid changes in pressure caused by the fluctuating flow rate and significant electromagnetic pickup from the pump, data is not collected during periods of recirculation.

Recirculation is aided by a resistive heater at the outlet of each chamber ensuring the evaporation rate is sufficient to maintain the desired flow rate. These heaters are coupled to a copper block surrounding the outlet tubing, all in a vacuum volume to thermally isolate this region from the Novec 7000. 

Recirculation is also used for the deployment of internal calibration sources, as described in \cref{sec:internal_sources}. In this method, forced flow of xenon over a radioactive source sweeps radon atoms into the test chamber. Additional ports along the recirculation manifold are flexible enough to allow testing of alternative purifying technologies for cross comparison. 

\section{Auxiliary Subsystems}\label{sec:monitoring}
\subsection{Slow Controls}
LabVIEW programs provide the main slow controls interface for both systems. Each system has its own National Instruments chassis through which all instrumentation for that system is controlled and read out. This includes:
\begin{itemize}
  \item Temperatures measured at 5--10 locations covering the test chamber, heat exchanger, and cryostat using T-type thermocouples and Pt100 RTDs
  \item Pressures measured at the test chamber inlet, outlet, in the Novec 7000 volume, and by a two-stage vacuum gauge in the pumping manifold
  \item Strain gauge mass measurements for the xenon storage bottle and Novec 7000 storage drum
  \item Mass flow measurements through the recirculation manifold
\end{itemize}
The LabVIEW programs also read and write data to and from the cryo chiller via RS232. Commands to turn on and off the cryo chiller, start the precooling sequence, and turn on or off cooling for a specified system can be sent using dedicated buttons on the LabVIEW front panel. The LabVIEW programs continually read the suction and discharge pressures in the cryo chiller along with the temperature at five locations. While the cryo chiller does have an internal interlock to shut down in the event of loss of cooling water flow, an additional flow switch is added externally to allow LabVIEW to redundantly handle and monitor this failure condition.

All data described above is written to a text file at a loop frequency that is typically set to be between 1 and 10 seconds. A Python-based data parsing class is used both for fast analysis of slow controls data by users, and by the monitoring system which regularly checks these parameters against a set of thresholds.

The temperature setpoint, electrode voltages, xenon pump, and data acquisition system are all controllable from the lab computers which can be accessed remotely. The only regular tasks that require the physical presence of an operator involve the operation of manual valves, typically used only during xenon filling and recovery. This ensures that between filling and recovery, the system can be operated entirely remotely.

\subsection{Monitoring Code}
The test facilities described here are designed to allow for operation by as few as 2 people. To enable that level of short-handed operation, a Python-based monitoring system was developed that automates the process of checking critical parameters and alerting operators. 
\begin{figure}[htbp]
  \centering
  \includegraphics[trim=40px 15px 40px 15px, clip, width=\textwidth]{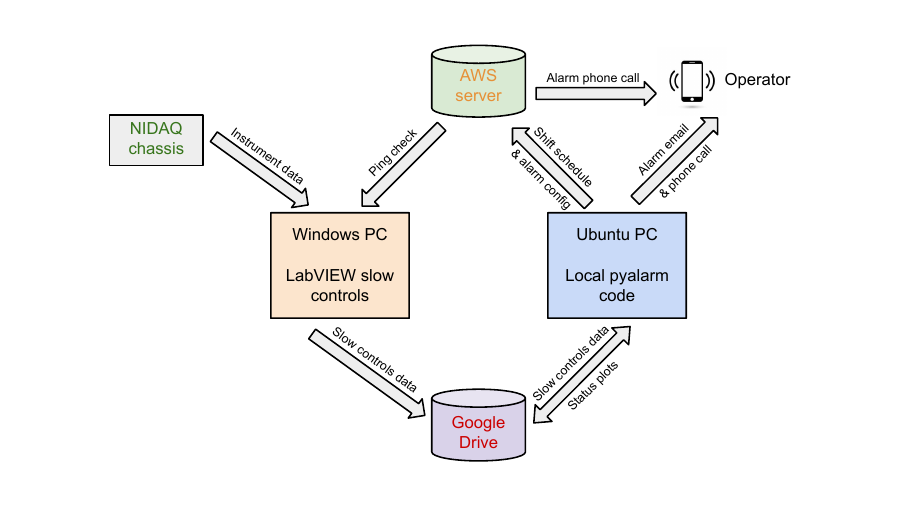}
  \caption{Functional schematic of the distributed monitoring system.}\label{fig:monitoring}
\end{figure}

\cref{fig:monitoring} shows a schematic of the monitoring system. The primary monitoring script runs on a desktop computer in the lab, while a second script runs on an off-site server managed by Amazon Web Services. Two configuration files may be updated as the monitoring program operates:
\begin{itemize}
    \item \textbf{Shift schedule}: a YAML file that contains a list of operators and a run coordinator, along with their email addresses, phone numbers, and start/end times for their shift. The monitoring code has been designed to ensure that there is always a designated operator that can be reached by phone.
    \item \textbf{Alarm config file}: a YAML file that contains allowable ranges for thermocouples, pressures, and any other parameter logged by the LabView system. If a parameter departs from its allowed range, the operator will be alerted by email and with regular phone calls until the issue is resolved.
\end{itemize}

The main loop of the monitoring code includes the following steps. The shift schedule and alarm configuration are loaded and synchronized with the remote monitoring server. The operator currently on shift is identified, and if they were not on shift previously, they are alerted with a phone call. Next, the most recent LabVIEW data file is downloaded from a Google Drive folder, the file is parsed, and the parameters are checked against the thresholds in the alarm config file. At the same time, a number of diagnostic plots are produced and uploaded to a Google Drive folder for quick interpretation by the operator. The monitoring loop will also ensure that the slow controls data is up to date and that the LabVIEW code is still writing current parameters as expected. If no errors occur, the script will sleep for the specified time until the next iteration (typically 5 minutes). 

Both the shift schedule and the alarm configuration can be modified at any time without interrupting the monitoring code. The entire code is wrapped in a loop with error-catching cases that call the operator upon encountering any unexpected problems to ensure that the monitoring never crashes without notifying. 

If an alarm is triggered by the monitoring code, almost all potential problems can be resolved remotely. All computers are remotely accessible, and a series of cameras placed around the lab allow for remote viewing of most hardware and instrumentation. Only errors originating from the cryo chiller, which require a manual reset, and power failures necessitate the physical presence of the operator.

In the event of a power outage in the lab, the remote monitoring server will detect that the local monitoring computer is no longer pingable and alert the operator. Some essential instruments in the lab are powered through an uninterruptible power supply (UPS) to ensure the xenon can be safely recovered without building power.

\subsection{Data Acquisition}
Both test stands have a dedicated data acquisition rack, distinct from the slow controls, which provides flexible triggering and digitization schemes. The most commonly used waveform digitizers are four Struck SIS3316 VME digitizers. Each unit consists of 16 channels with configurable 14-bit ADC dynamic range and sampling rate from 25 to 250~MS/s, allowing for simultaneous readout of multiple channels. These enable the development of the multi-channel charge and light sensors described below in \cref{sec:charge_light}.

\section{Experiments Currently Supported}
The test platforms described above are designed to be general-purpose systems capable of supporting a variety of R\&D efforts relevant to LXe detectors. A few of the xenon test chambers that are frequently operated within the Stanford xenon facility are described in this section, as well as the sensor technologies under development. Additional applications of the test stands are also discussed, including their use with specialized chambers for measurements of xenon purity and high voltage phenomena, and their application to the development of calibration techniques for xenon TPCs.

\subsection{Test Detectors}
\begin{figure}[htbp]
  \centering
  \includegraphics[width=\textwidth]{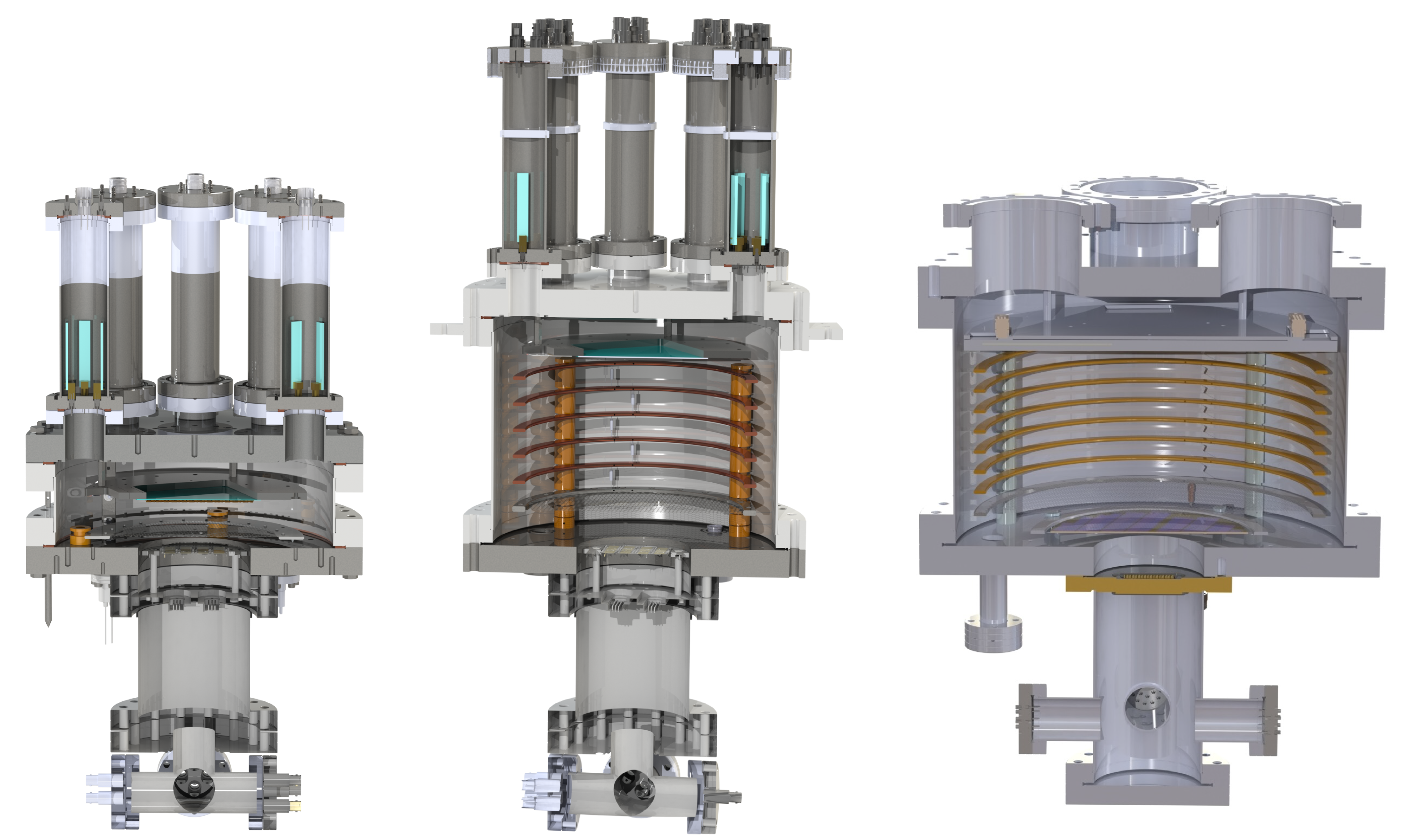}
  \caption{CAD renderings of the Short TPC (left), the Long TPC (middle) and the Wide TPC (right), at a 1:1 scale with one another.}
  \label{fig:tpcs}
\end{figure}

Three single-phase test TPCs (\cref{fig:tpcs}) are frequently used in the Stanford xenon facility. These detectors were designed primarily to test the prototype charge and light sensors for the nEXO experiment \cite{jewell_characterization_2018,dalmasson_large_2023} described in \cref{sec:charge_light}, and to develop calibration schemes, as described in \cref{sec:internal_sources}.

All three detectors have an active volume enclosed by a cathode grid at one end, a charge-sensitive anode plane at the other, and copper field-shaping rings spaced out along the length of the cylindrical drift region. The drift region is enclosed in a stainless steel CF spool piece with the charge-sensitive anode mounted to the top flange and a light sensor array mounted to the bottom flange beneath the cathode. The cathode grids are constructed from stainless steel etched hexagonal meshes with $\sim$95\% optical transparency. Four Vespel rods screwed to the bottom flange of the test chamber extend vertically upwards around the drift region. Regularly-spaced notches in these rods hold the cathode and field shaping rings. The TPCs and sensor arrays are fully modular; any test TPC can use any charge or light sensor array with the appropriate CF adapters.

\subsubsection*{Short TPC}
The ``Short TPC'' (STPC) is contained within a CF spool piece between two DN250CF flanges at either end. The drift region is 2.1~cm long, resulting in a drift time of $8~\mu \mathrm{s}$ at the nominal drift field of 380~V/cm \cite{albert_measurement_2017}. The STPC holds $\sim$10~kg xenon.

\subsubsection*{Long TPC}
Like the STPC, the ``Long TPC'' (LTPC) has a 19.3~cm diameter drift region between two DN250CF flanges. The LTPC has a significantly longer drift region of 13.8~cm, corresponding to a $\sim$~$70~\mathrm{\mu s}$ drift time at 380~V/cm. The drift region is surrounded by five field-shaping rings which keep the electric field uniform to within 80\% in the central 55\% fiducial volume. It holds $\sim$34~kg xenon.

\subsubsection*{Wide TPC}
The ``Wide TPC'' (WTPC) has a 16.1~cm drift length. It is contained within a DN350CF vessel and has a drift region with a diameter of 29.6~cm. With this larger diameter, the WTPC can hold $\sim$64~kg xenon. With 7 rings surrounding the drift region, an electric field uniformity of 95\% in the central 55\% fiducial volume is achieved. This TPC was designed specifically to test arrays of large area sensors which require the largest commercially available CF diameter. 

\subsection{Scintillation \& Ionization Sensors}
\label{sec:charge_light}
Scintillation light is collected by a square array of 24 FBK VUV-HD1 SiPMs grouped together in pairs, providing 12 light-collection channels. The SiPM array measures $6.5~\mathrm{cm}\times6.5~\mathrm{cm}$. It is mounted to a double-sided DN100CF flange, which can be bolted to the underside of the bottom flange of any test chamber. The bottom flanges of all test chambers have a 10.6 cm bore to accommodate the SiPM tile.

The small area of the SiPM tile compared to the test TPCs results in a low light collection efficiency. For this reason, the light signals are primarily used for triggering but not energy reconstruction. Alternatively, the flange carrying the SiPM array can be replaced with a UV quartz viewport and a UV photomultiplier tube.

A sub-D feedthrough routes SiPM signals from the LXe volume into a dry-nitrogen-filled volume containing preamplifier electronics \cite{fabris_2015}. This volume is sealed using CF to isolate the electronics from the Novec 7000. The preamplifier outputs are transmitted through feedthroughs to the Novec 7000 space, where RG58 cables transmit to the top flange and air space. 

The prototype ionization-charge detection module, or ``charge tile'', described in \cite{jewell_characterization_2018} has been used with both the LTPC and the STPC. In nEXO, charge tiles similar to this prototype will be arrayed to cover the entire anode plane, maximizing charge collection efficiency and single-site/multi-site discrimination \cite{li_simulation_2019}. 

One strength of a mid-scale LXe facility is in its ability to test arrays of prototype sensors, exploring potential unknown impacts on signal quality driven by inter-sensor interactions. The WTPC is specifically designed to explore a 2-by-2 array of $10~\mathrm{cm}\times10~\mathrm{cm}$ charge-tile modules, requiring a large chamber diameter. 

\subsection{Liquid Xenon Purity Monitor}
The xenon Purity Monitor (PM) (\cref{fig:compat_chambers}, left) consists of a 12 cm drift region, a gold-coated photocathode, an anode, and two Frisch grids. A charge signal is produced by the photoelectric effect using UV light injected through a 600 $\mu$m-core polyimide coated fiber in the xenon space. Light is sourced by a 60~W xenon flash lamp, liberating on the order of 100--1000 fC of charge per flash from the photocathode. A Frisch grid 1 cm from both the photocathode and anode isolate them from induction current while charges drift in the central region, and 14 field rings maintain a constant electric field of up to 400 V/cm in the central region. 

The electron lifetime is determined by the ratio of charge collected by the anode to charge liberated by the photocathode. Materials and test samples installed into two sample holders outside of the field cage allow for measurements of electronegative out-diffusion of elements to be included in large-scale xenon detectors. The design of the PM is based on a similar device described in \cite{carugno_electron_1990}.

\subsection{High-Voltage Test Chamber}
\begin{figure}[htbp]
  \centering
  \includegraphics[width=\textwidth]{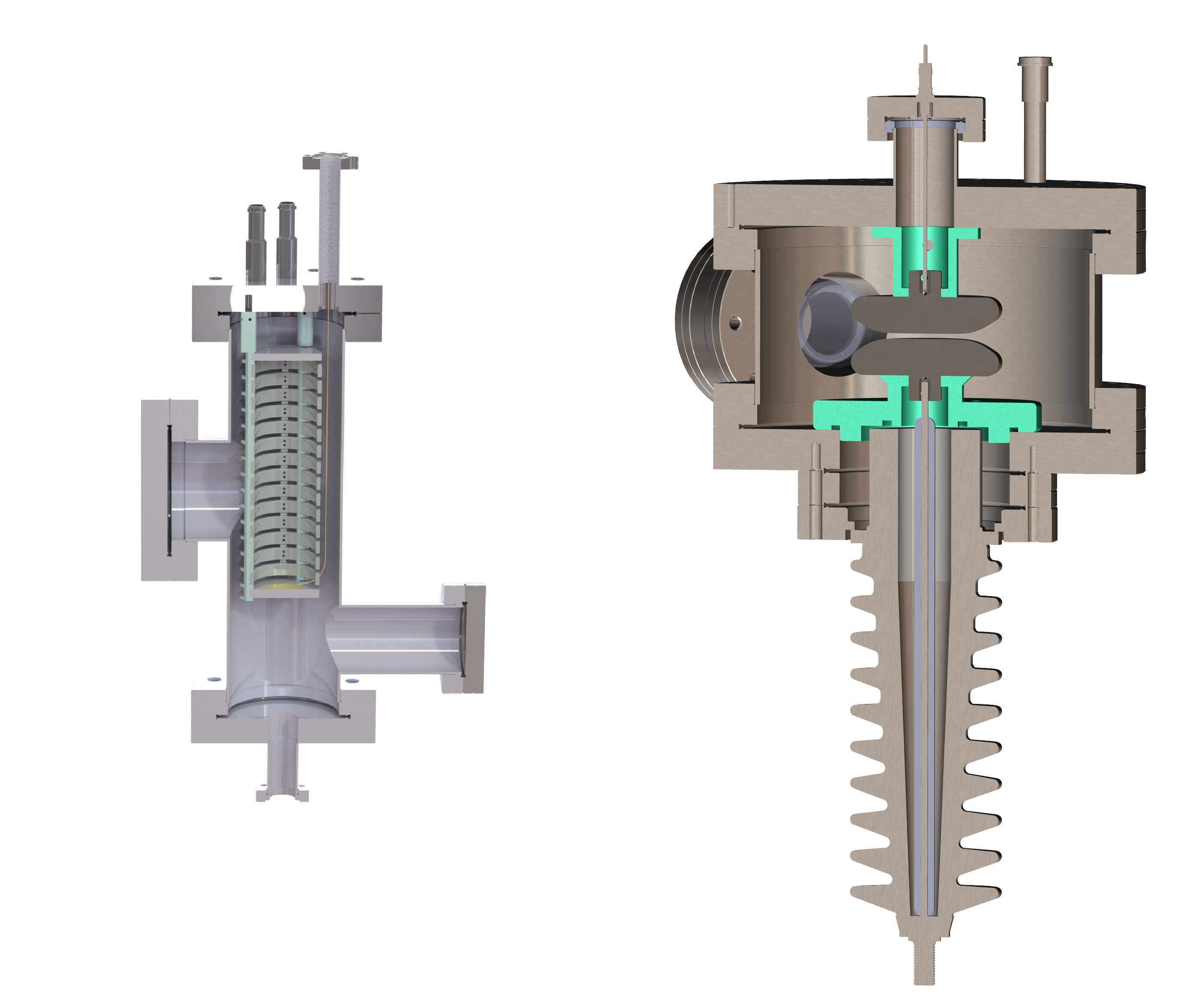}
  \caption{CAD renderings of the purity monitor (left) and high-voltage test chamber (right), at a 1:1 scale to one another.}
  \label{fig:compat_chambers}
\end{figure}

The High-Voltage Test Chamber (HVTC) (\cref{fig:compat_chambers}, right) is a 10~kg-scale LXe experiment that has observed a variety of high-voltage phenomena (HVP) using multiple pairs of polished electrodes with 16~cm$^2$ field-stressed area oriented in a plane-to-plane geometry with the ability to explore fields up to 60 kV/cm. The emphasis of the experiment is to explore the impact on HVP mitigation by depositing thin films of metals and insulators onto the surfaces of electrodes. A comparison of the performance of electrodes with surfaces of bare stainless steel, platinum, and aluminum-magnesium-fluoride has been explored. A publication detailing this experiment is in progress. 

Electrodes specifically shaped to create a large, uniform electric field in the central region are held at a fixed position, separated by 3.3~mm \cite{trinh_electrode_1980}. The stressed effective area of the surfaces of the electrodes is about 16~cm$^2$, and within that area the field is uniform to 3\%\footnote{Standard deviation over the surface of the electrode}. The stressed volume, a cylinder of radius equal to the stressed area contour, is 5.7~cm$^3$ with a field uniformity of 3\%. 

The cathode, attached to a 100~kV-rated high-voltage feedthrough, is biased at a ramp rate of 2~V/s. The anode is grounded through a passive integration circuit with a 400~$\mu$s time constant, and observes charge depositions across the gap at an equivalent noise level of 1~pC. Two 2.54~cm diameter photomultiplier tubes\footnote{Hamamatsu R8520-406} with a 90~degree azimuthal separation view the electrode gap through re-entrant, VUV-transparent vacuum windows. The photomultipliers record scintillation coincident with charge depositions, as well as cosmic rays.

\subsection{Calibration Systems}
Both systems allow for multiple types of calibration sources to be used, including internal $\alpha$ sources and internal and external $\gamma$ sources. The calibration sources are used both to develop calibration schemes for nEXO and to support tests of the sensor arrays described above. An overview of the deployment system for each source is outlined below.

\subsubsection*{Internal Sources}
\label{sec:internal_sources}
To calibrate the detector response for a particular test chamber, internal sources which generate events uniformly throughout the LXe volume are preferred. Dissolved $^{220}$Rn sources have been used to calibrate large-scale LXe TPCs \cite{ma_internal_2020,jorg_characterization_2023,lang_220rn_2016} by releasing $\alpha$ particles in the LXe volume. $^{127}$Xe has been shown to be a useful calibration source for low-energy events in LXe TPCs \cite{akerib_ultralow_2017,lenardo_development_2022}. The test stands described here have infrastructure to facilitate the deployment of both of these sources.

The $\alpha$-decays for calibration originate from an electrodeposited and diffusion-bonded oxide disc of $^{228}$Th from Eckert \& Ziegler\footnote{AF-228-PM at 18 kBq original activity.}. The disc is held in a CF enclosed volume equipped with particle filters on either end. The source volume is installed into the xenon recirculation path with a bypass valve and lockable radiation source valves.

The decay chain of $^{228}$Th includes a decay to $^{220}$Rn, which emanates from the surface of the disc into the flowing xenon gas. The $^{220}$Rn atoms are carried into the xenon chambers before they decay with a 55 second half life. Following a sufficiently long recirculation period, an equilibrium concentration of the 10.64 hour half life $^{212}$Pb is reached in the chamber. Short-lived daughters of $^{212}$Pb result in an $\alpha$-decay of $^{212}$Po which may be used for the electron lifetime and energy calibration. 

The internal $^{127}$Xe calibration source is produced through neutron activation on $^{126}$Xe by irradiating a sample cylinder of natural xenon, as described in \cite{lenardo_development_2022}. A 5.6~mL intermediate volume between the high-pressure sample cylinder and the recirculation manifold is used to inject quantities of the activated xenon into the recirculation path. The $^{127}$Xe mixes uniformly throughout the TPC, allowing uniform detector illumination with 408~keV electron recoil events which can be used for electron lifetime and energy calibrations. Due to the longer half life of 36.3 days, this calibration source is useful for repeated calibrations without having to frequently re-introduce the active isotope into the TPC.

\subsubsection*{External Sources}
External $\gamma$ sources are also useful as they allow for rapid deployment and removal. The Large System includes a removable source tube submerged within the Novec 7000 to allow external $\gamma$ sources to be inserted in close proximity to the test chamber while remaining isolated from the Novec 7000. The source, which is attached to the end of a plastic rod, can be positioned anywhere along the 80~cm length of the source tube. An Ultra-Torr fitting is used to create a seal around the plastic rod, isolating the air-filled interior of the source tube from the external air and preventing water vapor from being cryopumped into the cold interior. A sealed $^{232}$Th source is typically used in this system, with calibrations performed on events from the 2614~keV $^{208}$Tl $\gamma$ transition.

\section{Summary}
In this paper we have described a versatile test platform for the development of LXe TPC technology. It is specifically designed to operate LXe test chambers of up to $\sim$60~kg xenon mass for periods of weeks or more with as few as 2 remote operators. 

The use of an immersion-based cooling system is optimal for detectors at this scale and critical to accomplishing the operational goals above. The large thermal mass of cryogenic liquid stabilizes xenon temperatures in the case of a cooling failure, and liquid convection distributes heat evenly over chambers made of low-thermal-conductivity materials. 

The facility described in this paper is used to operate a variety of LXe test chambers that explore charge and light detection technologies, high-voltage phenomena, xenon purification, material out-diffusion, state-of-the-art cryogenic electronics, and cryogenic systems for end-to-end tests of prototype detector subsystems.

\acknowledgments
The authors thank Bob Conley of SLAC, whose welding services enabled the construction of many aspects of the systems described in this paper, Andrea Pocar, whose group at the University of Massachusetts Amherst provided the $^{220}$Rn calibration source, and Wesley Frey of the McClellan Nuclear Research Center, who helped with the development of the $^{127}$Xe calibration source. We are grateful to Jespere Calderone Nzobadila Ondze, Odwa Tyuka, Glenn Richardson, and Miao Yu for their assistance as remote operators of the xenon systems. We also gratefully acknowledge the support and advice of colleagues in the nEXO Collaboration. This work was funded by US DOE's Office of Science under grant DE-SC0017970.

\bibliographystyle{JHEP}
\bibliography{main}

\end{document}